\documentclass[twocolumn,showpacs, pre ,amsmath,amssymb]{revtex4}

\usepackage{graphicx}
\usepackage{dcolumn}
\usepackage{bm}

\usepackage[usenames,dvipsnames]{xcolor}
\definecolor{Highlight}{rgb}{1,1,0.75}
\usepackage[normalem]{ulem}					

\newcommand\ba{\begin{array}}
\newcommand\ea{\end{array}}
\newcommand\bp{\begin{picture}}
\newcommand\ep{\end{picture}}
\newcommand\be{\begin{equation}}
\newcommand\ee{\end{equation}}
\newcommand\bea{\begin{eqnarray}}
\newcommand\eea{\end{eqnarray}}
\newcommand\bs{\begin{subequations}}
\newcommand\es{\end{subequations}}
\newcommand\nn{\nonumber}
\newcommand\bfl{\begin{flushleft}}
\newcommand\efl{\end{flushleft}}
\newcommand\bsp{\begin{split}}
\newcommand\easp{\end{split}}

\newcommand\ri{\right}
\renewcommand\le{\left}

\newcommand{\tto}{\rightarrow}

\renewcommand\a{\alpha}
\renewcommand\b{\beta}

\renewcommand\d{\delta}

\newcommand\f{\phi}

\renewcommand\l{\lambda}

 \newcommand{\eqwithrates}[2]{\mathrel{\mathop{\rightleftharpoons}\limits^{#1}_{#2}}}

\begin{document}


\title{Mixed Poisson distributions in exact solutions of stochastic auto-regulation models}

\author{Srividya Iyer-Biswas}
\email[]{iyerbiswas@uchicago.edu}
\affiliation{James Franck Institute, University of Chicago, Chicago, IL 60637}
\author{C. Jayaprakash}
\affiliation{Department of Physics, The Ohio State University, Columbus, OH 43210}


\begin{abstract}
In this paper we study the interplay between stochastic gene expression and system design using simple stochastic models of auto-activation and auto-inhibition. Using the Poisson Representation, a technique whose particular usefulness in the context of non-linear gene regulation models we elucidate, we find exact results for these feedback models in the steady state. Further, we exploit this representation to analyze the  parameter spaces of each model, determine which dimensionless combinations of rates are the shape determinants for each distribution, and thus demarcate where in the parameter-space qualitatively different behaviors arise. These behaviors include power-law tailed distributions,  bimodal distributions and sub-Poisson distributions. We also show how these distribution shapes change when the strength of the feedback is tuned. Using our results, we reexamine how well the auto-inhibition and auto-activation models serve their conventionally assumed roles as paradigms for noise suppression and noise exploitation, respectively.
\end{abstract}

\maketitle




\section{Introduction}
Stochastic fluctuations in the numbers of key biochemicals may be significant compared to their mean levels. Such fluctuations arise even in a population of cells that were initially identical due to the inherently probabilistic nature of chemical reactions and the small numbers of reactants involved in key cellular processes like gene activation, transcription and translation~\cite{paulsson, vano}. Consequently, there may be significant cell-to-cell variability of gene products, in particular,  protein numbers. Stochastic gene expression and specifically, such  `intrinsic' fluctuations in protein numbers, have been the focus of several experimental and theoretical studies~\cite{raj, golding, raser, paulsson, vano, kepler, hornos, iyer-biswas, pedraza}. Biologically, the fluctuations in numbers of a given  protein could either be  desirable, or detrimental and thus requiring suppression, for the relevant cellular function~\cite{eldar}. An important systems biological goal is to understand how the noise characteristics associated with a given gene regulatory network inform the biological function of the corresponding protein.

Previous studies have indicated that tight control of protein numbers, when desired, is often achieved by an auto-repression motif of gene expression~\cite{savageau, becskei00, thattai, dublanche, austin}. It has even been argued that the reason why this network motif occurs far more frequently in nature (40\%  of the known transcription factors in {\em E. Coli} are controlled by negative auto-regulation~\cite{thieffry}) than in studies of randomized networks is because it achieves stability against fluctuations~\cite{alon}.  On the other hand, bimodal distributions of protein numbers may be exploited by cells to dynamically switch between different expression states; this is especially useful for cellular processes where conditional `locking' of sub-populations of cells into distinct fates needs to be achieved without changing the underlying network structure. The auto-activation motif has been implicated in  systems in which such tunable population heterogeneity is desirable~\cite{becskei01, acar08, smits, suel}.

Here we explore the interplay between stochastic gene expression and systems design by examining two simple stochastic gene pulsing models with auto-regulation. It is now well established that in many genes  transcribe and/or translate in `bursts', i.e., mRNA/proteins are produced with significantly varying dead times in between successive rounds of production~\cite{raj, 2005-paulsson-eu, iyer-biswas,  2006-friedman-zl, vano}. This important aspect of gene expression is encapsulated in a model in which the gene can stochastically switch between long-lived ``off" states and ``on" states leading to intermittent mRNA and protein expression. {(We prefer the term ``pulsing'' to ``bursting'' since the latter terminology could be misleading~\cite{2005-paulsson-eu}.)} Therefore, in the positive (negative) feedback model considered here, the amount of protein produced is assumed to proportionally increase the propensity of the gene to dwell in the on (off) states.

We use the Poisson representation, first introduced in~\cite{1977-gardiner-kq}, a technique whose particular usefulness in the context of analyzing feedback models of gene expression we elucidate here. While models of stochastic gene expression, including those for auto-regulation, have been previously considered~\cite{2005-paulsson-eu, hornos, iyer-biswas,  2006-friedman-zl, 2008-visco-qv, 2009-visco-sf}, what has been lacking is a systematic prescription for classifying where in the multi-dimensional parameter space of each model qualitatively distinct distributions are obtained. Typical solutions and examinations of these models utilize the generating function method for solving the corresponding Master equation, or numerical simulations based on the exact Gillespie algorithm, or combinations thereof. However, in both these approaches, systematically classifying the entire parameter space of the model is not feasible in general. Here we show  how such a classification is possible when the physically well-motivated Poisson representation is instead used, since it naturally yields the particular dimensionless combinations of parameters that are the important ones for each model. We use this representation to both derive the exact steady state protein distributions in the two models considered here, and to also analyze the respective parameter spaces, demarcating where in them bimodal, power-law tailed, sub-Poisson and other distributions occur.  Using the classification of allowed distributions, we then re-examine how well the models of negative and positive feedback considered here serve their conventional roles as paradigms for noise suppression and noise exploitation, respectively. 

While the idea of writing down protein distributions as exact linear superpositions of Poisson distributions is relatively new~\cite{iyer-biswas}, mixtures of  Poisson distributions have  been long studied in various contexts including photon statistics in quantum optics~\cite{sudarshan} and in accident proneness models in actuarial sciences~\cite{grandell}. Remarkably, in both the auto-activation and the auto-repression cases, we find classes of mixed Poisson distributions that, to the best of our knowledge, have not been previously considered~\cite{karlis}. Moreover, they arise {\em dynamically} in these models. We also show that the Beta-Poisson mixture, which has been previously utilized as a  versatile {prior} distribution in accident proneness models~\cite{grandell}, naturally arises as a limiting case from these dynamics.
\section{Theoretical framework}
\subsection*{The Poisson representation}
Detailed expositions of the Poisson Representation can be found in~\cite{1977-gardiner-kq, gardiner, grandell}. We have briefly discussed application of the Poisson Representation to linear models of gene regulation without feedback in~\cite{iyer-biswas}.  Here analyze how the exact steady-state protein distributions $P(p)$ of the models with  feedback, positive or negative, may be represented as a superposition of Poisson distributions, with a weighting probability density $\rho(\l)$ for the Poisson mean $\l$. In other words, we determine whether a probability density $\rho(\l)$ can be found such that
\begin{align}
P(p) = \int_{0}^{\infty}  \frac{e^{-\l} \,\l^{p}}{p!} \rho(\l) \,d \l.
\end{align}
If indeed such a probability density $\rho(\l)$ can be found, an immediate implication is that the corresponding $P(p)$ must be super-Poisson, i.e., its variance must be greater than its mean, and thus the ratio of the two, the Fano-factor (FF), must be greater than unity. (In contrast, the Poisson distribution has variance equal to mean and thus an FF $=1$.)

The superposing  or mixing density, $\rho(\l)$, is a function of a continuous variable, $\l$.  In contrast, $P(p)$ is a function of $p$, which is only allowed discrete (positive integer) values. Thus the convexity and monotonicity properties of $\rho(\l)$ are easier to ascertain than that for $P(p)$. In turn, these properties determine the allowed shapes of $P(p)$ for a given stochastic gene expression model. Specifically, bimodal $P(p)$ distributions correspond to concave (upwards) $\rho(\lambda)$; power-law tails in $P(p)$
arise when $\rho(\l)$ itself has a monotonically decreasing power law tail;  a monotonically increasing $\rho(\lambda)$ leads to a unimodal $P(p)$ distribution with the mode approximately at the upper edge of the $\l$ interval; when $\rho(\l)$ is concave downwards with  a maximum at some intermediate value of $\l$, then unimodal $P(p)$ distributions with a mode around the same value result. We use the exact, analytical expressions that we derive for  $\rho(\l)$ to map out where in the parameter space each qualitatively distinct shape of $P(p)$ arises.

\subsection*{Master equations for the auto-activation and auto-repression models}

{\bf Auto-activation.} The auto-activation model considered here is given by the following reactions, with the protein switching the gene from the off to the on state:
\begin{align}
D &\eqwithrates{c_f}{c_b} D^*, \nn \\
D+P &\stackrel{a}{\longrightarrow} D^*+P, \nn \\
D^*&\stackrel{p_b}{\longrightarrow} D^*\,+\,P,  \nn \\
P &\stackrel{p_d}{\longrightarrow} \emptyset \,.
\end{align}
We use  $P_0(p,t)$ and $P_1(p,t)$ to denote the probabilities that there are $p$ proteins at time $t$ and that the gene is in the off and on state respectively.  The Master equations for the time-evolution of these probabilities are then obtained using standard techniques~\cite{gardiner}. They are
\begin{align}
\label{eq-aa-p-ME}
\frac{dP_0(p,t)}{d t}&= -c_fP_0(p,t)\,+\,c_bP_1(p,t)\,-\,a p P_0(p,t)  \nn
\\
&+\,p_d\,[(p+1)P_0(p+1,t)-p P_0(p,t)\,],  \nn \\ 
\frac{d P_1(p,t)}{dt}&= c_fP_0(p,t)\,-\,c_bP_1(p,t)\,+\,a p P_0(p,t)\,\nn
\\
&+\,p_d\,[(p+1)P_1(p+1,t)-p P_1(p,t)\,]  \,\nonumber
\\
&+p_b\,[\,P_1(p-1,t)\,-\,P_1(p,t)\,].
\end{align}
We define $\rho_{0}(\l)$ and $\rho_{1}(\l)$ as
\begin{align}
P_\alpha(p,t)\,\equiv\,\int_0^\infty d\lambda\,\rho_\alpha(\lambda,t)\,e^{-\lambda}\,\frac{\lambda^p}{p!} \mbox{ for } \a = 0 \mbox{ or } 1,
\end{align}
and note that $\rho(\lambda)\,= \rho_{0}(\l) + \rho_{1}(\l)$ satisfies the normalization condition $\int d\lambda\,\rho(\lambda)\,=\,1$. The corresponding Master equations for $\rho_\alpha(\l)$ are then given by 
\begin{align}
\label{eq-aa-rho-ME}
\partial_t\rho_0(\lambda,t)&= -c_f\rho_0(\lambda,t)\,+\,c_b\rho_1(\lambda,t)\,+\,\,
\partial_\lambda[\lambda\rho_0(\lambda,t)] \nonumber \\
& -a \left(\lambda\rho_0\,-\, \partial_\lambda(\lambda\rho_0)\right) \nn \\
\partial_t\rho_1(\lambda,t)&= c_f\rho_0(\lambda,t)\,-\,c_b\rho_1(\lambda,t)\,+\,\,
\partial_\lambda[\lambda\rho_1(\lambda,t)]\nonumber  \\
&+\,a \left(\lambda\rho_0\,-\,\partial_\lambda(\lambda\rho_0)\right)\,-\,p_b\partial_\lambda\rho_1(\lambda,t)\,,
\end{align}
with the boundary condition
\begin{align}
\pm \,a\,\left. e^{-\lambda}\,\frac{\lambda^p}{p!}\,\lambda\rho_0(\lambda)\,\right\vert_0^{\lambda_{max}}\, = 0,
\end{align}
for $0 \leq \l \leq \l_{max}$; $\l_{max}$  needs to be computed.  In going from Eq~\ref{eq-aa-p-ME} to Eq~\ref{eq-aa-rho-ME}, we have imposed the condition that the boundary terms resulting from integration by parts vanish.  The solution we obtain does indeed behave as required and so the assumption that the boundary terms vanish can be justified a posteriori (see Results section).

{\bf Auto-repression.} The auto-repression model considered here is given by the reactions
\begin{align}
D&\eqwithrates{c_f}{c_b} D^*, \nn \\
D^*+P&\stackrel{r}{\longrightarrow} D+P, \nn \\
D^*&\stackrel{p_b}{\longrightarrow}D^*\,+\,P,  \nn \\
P &\stackrel{p_d}{\longrightarrow}\emptyset \,.
\end{align}
We can derive the Master Equations satisfied by the $\lambda$-densities as before.
\begin{align}
\label{eq-ar-rho-ME}
\partial_t\rho_0(\lambda,t)&= -c_f\rho_0(\lambda,t)\,+\,c_b\rho_1(\lambda,t)\,+\,\,
\partial_\lambda[\lambda\rho_0(\lambda,t)] \nonumber \\
& +\,r\,\left(\lambda\rho_1\,-\, \partial_\lambda(\lambda\rho_1)\right) \nn \\
\partial_t\rho_1(\lambda,t)&= c_f\rho_0(\lambda,t)\,-\,c_b\rho_1(\lambda,t)\,+\,\,
\partial_\lambda[\lambda\rho_1(\lambda,t)] \nonumber \\
&-\,r\,\left(\lambda\rho_1\,-\,\partial_\lambda(\lambda\rho_1)\right)\,-\,p_b\partial_\lambda\rho_1(\lambda,t)\,,
\end{align}
with the boundary condition
\begin{align}
\,-\left. e^{-\lambda}\,(p_b-\lambda-r\lambda)\,\frac{\lambda^p}{p!}\,\rho_1(\lambda)\,
\right\vert_0^{\lambda_{max}} =0
\end{align}
 for $0 \leq \l \leq \l_{max}$, where such a $\l_{max}$ must be found.

\begin{figure}
\begin{center}
\rotatebox{-90}{\resizebox{7.2cm}{!}{\includegraphics{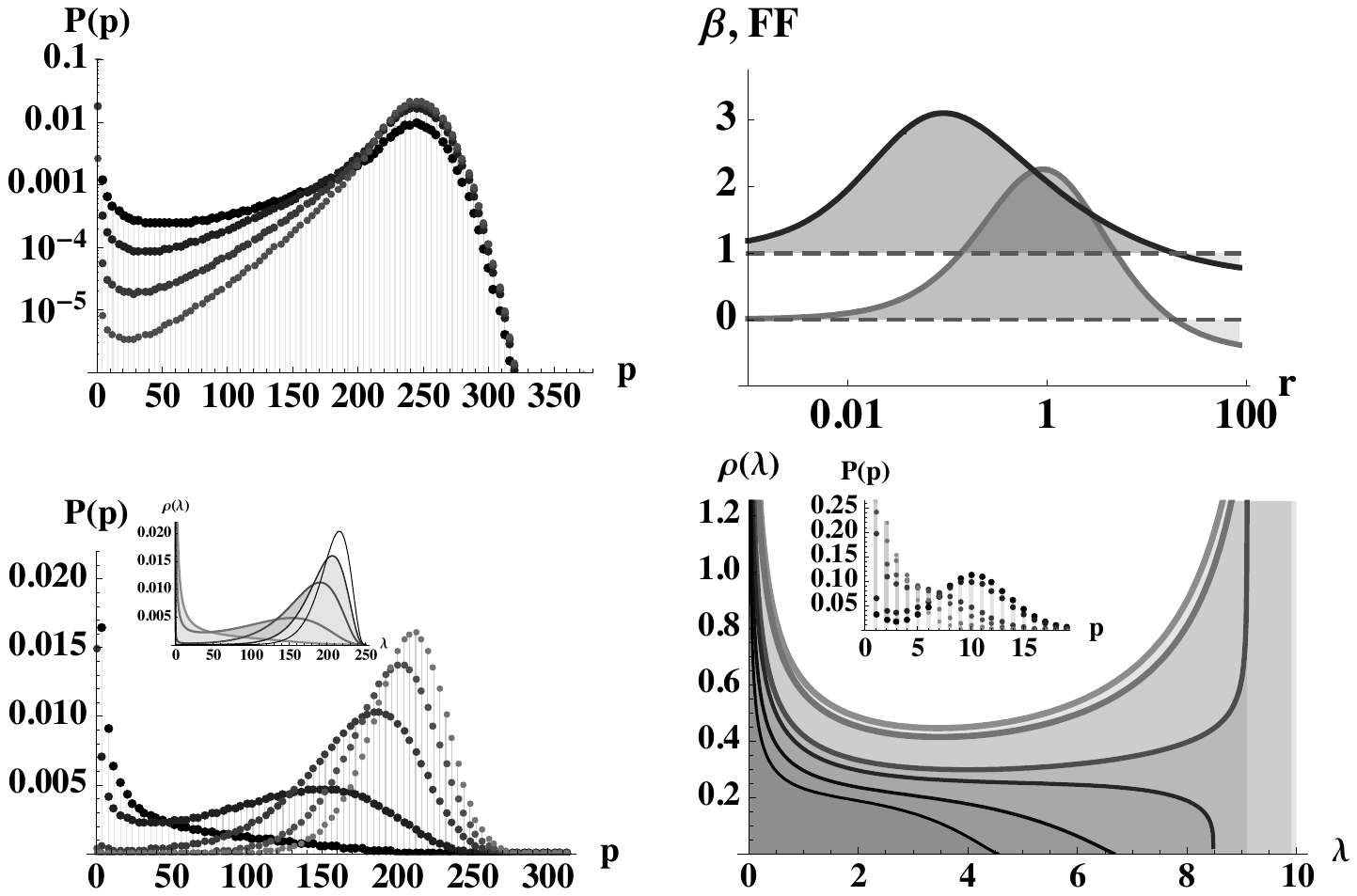}}}
\caption{{\em Top Left}:  The response of the protein distribution to increasing activation strength in the  $\f <1, \b<1$ quadrant in the auto-activation model resembles the classic `binary response' associated with auto-activation systems. {\em Bottom Left}: The effect of increasing activation strength, $a$, on the protein distribution in the fourth quadrant ($\f <1, \b>1$)  is `graded'.  {\em Top Right}: The effect of increasing auto-repression strength, $r$, on the Fano-factor of the protein distribution in the auto-repression model.  For the values chosen, both $\b$ and the Fano-factor go through maximum values at (different) intermediate values of $r$, before the distribution becomes sub-Poisson after the threshold value $r = r_{0}$; this happens exactly when $\b = 0$. {\em Bottom Right}: Effect of increasing auto-repression strength $r$, on the protein distribution in the auto-repression model. Six different points from the above figure are chosen from the range where $\b$ remains positive. The other rates are the same as the ones used in the previous figure.  $r$ increases from lighter to darker values.}
\label{fig-2}
\end{center}
\end{figure}

\section{Results}

To place our results for the auto-activation and auto-repression models in context, we will find it useful to compare these results with those derived for the linear pulsing model in \cite{iyer-biswas}. Both auto-regulation models reduce to the linear pulsing model (LPM) in the limit where the auto-activation strength, $a$, or the auto-repression strength, $r$, tends to $0$. In ~\cite{iyer-biswas} we have also shown how the `phase-diagram' of all possible distributions for the LPM can be classified in terms of the two rescaled dimensionless rates $c_{f}/p_{d}$ and $c_{b}/p_{d}$.
\subsection{Auto-activation}

The coupled Master equations,  Eq~\ref{eq-aa-rho-ME}, can be solved using standard techniques and give
\begin{align}
\rho(\lambda)\, = \mathcal{N} \,e^{\frac{a}{p_{d}+a}\lambda}\,\,\lambda^{\frac{c_f}{p_{d}+a}-1}\,\, \le(\frac{p_{b}}{p_{d}}-\lambda \ri)^{\frac{c_b}{p_{d}+a}\,-\,1}
\end{align}
with $0\leq \l \leq p_{b}/p_{d}$; $\mathcal{N}$ is the normalization constant. The choice of $\lambda_{max} = p_{b}/p_{d}$ ensures that the boundary terms vanish, as required.
This exact expression leads naturally to the correct parametrization of the combinations of the rate constants that are relevant for analyzing this nonlinear model. We  rescale $\l$ by $p_{b}/p_d$ so that it lies between $0$ and $1$. It is useful to rescale all rates by the effective protein degradation rate, $p_{d}$. For convenience in classifying the different kinds of protein distributions that arise in this model, we define the following parameters:  $\a  \equiv a\,p_{b}/(1+a), \, \f \equiv c_{f}/(1+a)$ and $\b \equiv  c_{b}/(1+a)$.
 We then have
\begin{align}
\rho(\lambda)\,= \mathcal{N}e^{\a\,\lambda} \,\,\lambda^{\f-1}\,\, (1-\lambda)^{\b-1}.
\end{align}
Note that $\f$ and $\b$ characterize the singularity at the upper and lower limits of $\l$. 
 Using this superposition-of-Poissons representation we have found that in each of the four `quadrants' determined by $\f$ and $\b$ greater or less than $1$, the protein distribution has a distinct shape. Since the superposing density, $\rho(\l)$, is found to extend from $\l=0$ to $\l = p_{b}$, $P(p)$ extends till $\sim p_{b}$. When the density diverges at both limits, i.e., $\f$ and $\b < 1$ yielding a  $\rho$ that is concave upwards the protein distribution is bimodal. When $\rho$ vanishes at both limits, i.e., $\f$ and $\b > 1$, yielding a $\rho$ that is concave downwards a broad bell-shaped distribution of proteins arises.

As the autoactivation strength $a \tto 0$, $\a \tto 0$, $\f \tto c_{f}$ and $\b \tto c_{b}$ the protein distribution of the autoactivation model becomes the exact steady-state distribution~\cite{iyer-biswas} obtained in the LPM. The latter is  a Beta distribution
\begin{align}
\rho(\lambda)\,= \mathcal{N}\,\,\lambda^{c_{f}-1}\,\, (1-\lambda)^{c_{b}-1}.
\end{align}
Thus, the `phase-diagram' of possible distributions in this model is very similar, in large regions of the parameter space $\f$ and $\b$,  to that of the LPM, despite the auto-activation, once we identify $\f$ and $\b$ in this model with $c_{f}$ and $c_{b}$ in the simple pulsing model.

We focus on the most interesting new feature that arises in this `phase-diagram' in this model. Consider the quadrant where $\f<1$ and $\b>1$. When $a=0$, i.e., in the LPM, we have found ~\cite{iyer-biswas} long-tailed distributions with power-law behavior. In the auto-activation model, in contrast, two possibilities arise  depending on whether $\a$ is lesser or greater than $\a_c\,\equiv\,(\sqrt{1-\f}+\sqrt{\b-1}\,)^{2}$. In the former case long-tailed distributions with power-law regions arise, with an exponent $\f-1$  as in the $a=0$ case.

For $\a\,>\,\a_c$  the distribution becomes an unusually behaved {\em bimodal} distribution! To appreciate its nature we recall that when both $\f$ and $\b$ are $<1$ (Figure~\ref{fig-2}) bimodal distributions occur with the two modes always at $0$ and $p_{b}$, i.e., at the edges of the allowed values of $\l$.  As $a$ the activation strength increases,  the weights around $0$ and $p_{b}$, are redistributed without affecting the separation between the modes. This is the classic `binary response' ~\cite{becskei01} typically associated with auto-activation: cells may be thought to be divided into two sub-populations with low and high protein numbers and increasing activation strength only changes their relative proportions. In contrast, the new bimodal distribution exhibits a second mode not  at $p_{b}$, the maximum allowed value of $\l$, but at intermediate values.  As $\a$ is increased by increasing $a$, the protein distribution goes from being monotonically decreasing power-law to bimodal because auto-activation affects cells with intermediate numbers of proteins the most. Thus, when the feedback strength is strong enough that $\alpha>\alpha_c$, a new minimum and as well as a new maximum develop in $\rho(\l)$, at intermediate values of $\l$. Correspondingly, $P(p)$ becomes bimodal with the second mode arising at a value of $p < p_{b}$.   As the activation strength increases, this mode tends to higher values of $p$,  but the weight at $0$ (the first mode) simultaneously erodes rapidly making the distribution effectively unimodal for strong enough activation. Thus in this quadrant even though bimodal distributions arise for intermediate activation strength, the response to increasing activation is really `graded' as illustrated in Figure~\ref{fig-2}. As $a$  increases, the protein distribution goes from being negatively skewed, with a large likelihood of obtaining a small number of proteins to a positively skewed distribution, with a large likelihood of obtaining a large number of proteins.

We point out the possible relevance of our results to the observation in a recent experiment of  Maheshri et al.~\cite{maheshri} of bimodal protein expression in a synthetic
yeast system with positive feedback and no cooperativity as in our model. As the activation  increases,
their distribution goes from a broad bell-shaped distribution to the bimodal distribution similar to the one described above.  Our model explains their observation of graded response of the 1xtetO promoter with increasing auto activation strength. As expected from our model, with increasing $a$,
the mode at larger value travels further towards the right and acquires more weight
until a Poisson like distribution occurs.

Since this model is nonlinear the equations for all the moments are coupled and one needs the full distribution to obtain even the lowest two moments. 
Using the exact solution for the distribution, one can  evaluate the  Fano-factor (FF),  the ratio of the variance to the mean of a distribution. The FF may or may not go through a maximum value as $a$ is increased, but beyond a threshold FF always decreases with $a$ and tends to $1$ as $a \tto \infty$.  Thus increasing auto-activation results in noise reduction, a role not conventionally associated with positive feedback.  This is true since  the gene is always ``on'' as the  activation strength tends to infinity, and a Poisson protein distribution, with $FF = 1$ results. For any initial choice of parameters, for large enough $a$, both $\f$ and $\b$ fall below $1$, and the protein distribution becomes bimodal. However, in the limit $a \tto \infty$, the mode at $0$ is entirely eroded and $\rho(\lambda) \tto \d(p_{b}-\l)$: $P(p)$ becomes Poisson.


\subsection{Auto-repression}
The analysis of this model proceeds along the same lines as the auto-activation model. Once again, this formulation leads naturally to the correct parametrization of the combinations of the rate constants that are relevant for analyzing this nonlinear model. We define the new parameters, $\a \equiv r\,p_{b}/(1+r)^{2}, \, \f \equiv c_{f}$ and $\b \equiv  p_{b}\,r/(1+r)^{2} + (c_{b}-c_{f}\,r)/(1+r)$. All rates have been scaled by the protein degradation rate, $p_{d}$ as before. In terms of these new variables, the steady state generating function is identical in form to that derived in the auto-activation model!

However, there is a subtle difference which has profound consequences: the parameter $\b$ can become {\em negative} for suitably chosen rates, $p_{b}, c_{f}, c_{b}$ and $r$, in this model, unlike the auto-activation case. Thus the weighting probability density $\rho(\l)$  can be found, only if $\b >0$. This immediately implies that for $\b >0$, the protein distribution in the auto-repression model is super-Poisson, i.e., its FF is $>1$ and thus `noisier' than the Poisson distribution that arises in the simple birth-death model.

When $\b$ is $ < 0$, we find that the protein distribution becomes sub-Poisson, i.e., its FFs becomes $< 1$. Thus, only when $\b<0$ can the auto-repression be said to be strong enough to cause reduction of the noise level in related models, such as  the LPM and the auto-activation model.  On analyzing the condition $\b<0$, we find that for any given value of the rates $c_{f}, c_{b}$ and $p_{b}$, there is a threshold value of the repression strength, $r_{0}$, such that when $r$ increases beyond this threshold value, the distribution becomes sub-Poisson (as illustrated in Figure~\ref{fig-2}). As seen in the top right panel of Figure~\ref{fig-2} this suppression occurs for large values of $r$ and over a narrow range. Exactly at the threshold value, the Fano-factor is found to be unity. The expression for $r_{0}$ is
\begin{align}
r_{0} = \frac{c_{b}+p_{b}-c_{f}+\sqrt{(c_{b}+p_{b}-c_{f})^{2}+ 4\,c_{f}\,c_{b}}}{2\,c_{f}}
\end{align}
For values of $r >r_{0}$, i.e., when the distribution is sub-Poisson, a formal expression for $\rho(\l)$ may be derived, with the understanding that it can no longer be interpreted as a probability density. In fact, $\l$ now extends over the complex plane. Remarkably, even in this case, the functional form of $\rho(\l)$ remains the same for a suitably chosen contour.  In this case, depending on whether $c_{f}$ is $< 1$ or $> 1$, the protein distribution is a monotonically decreasing or a sharply peaked bell-shaped distribution, respectively.

When $\beta >0$, we find that the different possible distributions of the auto-activation models all occur for auto-repression for appropriate values of $\alpha, \beta$ and $\nu$, when $\beta >0$, as illustrated in Figure~\ref{fig-2}. This underscores the inadvisability of  naively inferring that the choice of auto-inhibition motif is designed to obtain noise suppression without further exploring the specific details of the system. See also ~\cite{paulsson2010} for a control and information theoretical perspective on the issue. Quantitatively,  $\lambda$  is in the range $ 0 \leq \lambda \leq p_{b}/(1+r)$ and so the protein distribution extends to about $p \sim p_{b}/(1+r)$.  The effective parameter $\beta$ is now a function of all the rates in the problem while the effective parameter $\phi\,=\,c_f$ as in  the linear pulsing model.
\section{Concluding Remarks}
The auto-regulation motif  is ubiquitous in gene regulation~\cite{alon, vano}. The auto-regulation models studied here are admittedly simplified descriptions of those observed in nature:  we have not included separate transcription and translation steps. In prokaryotes, since mRNAs are rapidly translated into proteins, this is typically a reasonable approximation. For eukaryotic systems, when the mRNA time-scale is significant, these results should not be applied literally. Further, the effects of co-operative auto-regulation are not included in our models. However, even in this simple model a plethora of behaviors are observed including power laws, and bimodal distributions that behave in a graded fashion and sub-Poisson statistics.
We have also established the utility of the Poisson representation which yields quite naturally, the important, scaled, dimensionless parameters that characterize  non-linear gene regulation models. 
We have shown that auto-activation produces `binary' responses to increasing activation strength and that auto-repression produces noise-suppressed sub-Poisson protein distributions in very limited regions of the parameter space. Our work serves to add a note of caution to assuming that positive and negative feedback, when found in natural biological systems, are present to serve these purposes.

\section{Acknowledgements}
CJ acknowledges support through contract HHSN272201000054C of NIAID. SIB  acknowledges  useful discussions  with N. Maheshri, A. Walczak, N. Wingreen and R. R. Biswas. 



\end{document}